\documentstyle[epsfig]{elsart}
\newcommand{\gsim}{\mathrel{\rlap{\lower4pt\hbox{\hskip1pt$\sim$}}
\raise1pt\hbox{$>$}}}
\newcommand{\lsim}{\mathrel{\rlap{\lower4pt\hbox{\hskip1pt$\sim$}}
\raise1pt\hbox{$<$}}}

\begin{document}
\begin{frontmatter}
%
\title{The $\gamma^*\pi^0\to\gamma$ form factor }
\author[dubna]{M. A. Ivanov} and
\author[dubna,tomsk]{V. E. Lyubovitskij}
\address[dubna]{Bogoliubov Laboratory of Theoretical Physics, \\
Joint Institute for Nuclear Research, 141980 Dubna, Russia}
\address[tomsk]{Department of Physics, Tomsk State University,\\
634050 Tomsk, Russia}
%
\begin{abstract}
The $\gamma^*\pi^0\to\gamma$ form factor is obtained within
the Lagrangian quark model with separable interaction
known to provide a good description of the pion observables at low energies.
The pion-quarks vertex is chosen in a Gaussian form.
The form factor obtained is close to the available experimental data
and reaches smoothly the Brodsky-Lepage limit at $Q^2\approx 10$ GeV$^2$.
\end{abstract}
\end{frontmatter}
%
\section{Introduction}
\baselineskip 20pt
The study of neutral pseudoscalar mesons in two-photon reactions
provides information on their electromagnetic structure. While the
magnitude of $\pi^0\to\gamma\gamma$ amplitude is defined by the
Adler-Bell-Jackiw axial anomaly ~\cite{ABJ} which maybe derived from
the simple quark diagram with local pion-quark interaction, the knowledge
of the form factor with one photon being off mass shell requires
the introduction of a nontrivial distribution of quarks inside the pion.
The behavior of the off-shell axial anomaly is studied in the
$\gamma^*\pi^0\to\gamma$ reaction. The data available for this transition
form factor in the space-like region $Q^2<2.2$ GeV$^2$ are from
the CELLO Collab. \cite{CELLO}. New data covering the $Q^2$ region from
2 GeV$^2$ up to 20 GeV$^2$ have been reported by the CLEO Collab.
\cite{CLEO}. There is a project to measure this form factor via virtual
Compton scattering from a proton target at Jefferson Lab. \cite{CEBAF}.

The $\gamma^*\pi^0\to\gamma$ transition form factor has been studied
in the models based on triangle diagrams with nontrivial pion-quark
vertex and local quark propagators\cite{Anikin,Ito}, and
within a QCD-based model which implements dressing of the vertices and
propagators consistent with quark confinement~\cite{Roberts}.
This transition is a clean test of QCD predictions for exclusive processes.
The leading asymptotic behavior $F_{\gamma\pi}(Q^2)\to
4\pi^2f_\pi^2/Q^2$ was obtained in~\cite{Brodsky}. QCD sum rules
were applied to calculate this form factor in the region $Q^2>1$ GeV$^2$
\cite{Rad}. A hard scattering approach which includes transverse
momentum effects and Sudakov corrections was developed in \cite{Kroll}
to describe the large momentum transfer behavior of meson-photon
transition form factors.

In this brief note we calculate the $\gamma^*\pi^0\to\gamma$ transition form
factor within the approach developed in \cite{Anikin}. For the first time,
we derive the representation for this form factor from the triangle diagram
with dressed pion-quark vertices described by an arbitrary function
decreasing rapidly in the Euclidean region. This representation is
valid for any momenta of the pion and the photons. We use a Gaussian form for
the vertex to make the numerical calculations. The model parameters
(the cutoff parameter $\Lambda=1$ GeV and the constituent quark mass
$m_q=237$ MeV) are  determined from the $\pi$-decay constants and
provide a good description of the pion observables at low energies.
We find that the form factor obtained is close to the available experimental
data  and reaches smoothly the Brodsky-Lepage limit at
$Q^2\approx 10$~GeV$^2$.

\section{The $\pi^0\gamma\gamma$ vertex function}
\baselineskip 20pt

The basis of the approach \cite{Anikin} is the interaction Lagrangian
describing the transition of hadrons into constituent quarks
and {\it vice versa} via an effective vertex function related to
the momentum distribution of the constituents. For our purpose,
we write the pion Lagrangian as

\begin{eqnarray}\label{lagran}
{\cal L}_\pi^{\rm int}(x)&=&i\frac{g_\pi}{\sqrt{2}}{\bf\pi}(x)
\int\hspace*{-.1cm}dy_1\int\hspace*{-.1cm}dy_2 f([y_1-y_2]^2)
\delta\left(x-{y_1+y_2\over 2}\right)\bar q(y_1)\gamma^5
{\bf\tau}q(y_2)
\nonumber\\
& &\\
&=&i\frac{g_\pi}{\sqrt{2}}{\bf\pi}(x)
\int\hspace*{-.1cm}dy f(y^2)\bar q(x+y/2)\gamma^5 {\bf\tau}q(x-y/2).
\nonumber
\end{eqnarray}
The vertex function $f([y_1-y_2]^2)$ depending upon the relative coordinate
only is chosen to garantee ultraviolet convergence of the Feynman diagrams.
At the same time this function is a phenomenological description of the
long-distance QCD interactions between quarks and gluons.
The Fourier transform of the vertex function in Minkowsky space
is defined as $f(y^2)=\int d^4k/(2\pi)^4 f(k^2)$.
The gauging of the nonlocal Lagrangian Eq.~(\ref{lagran}) has been done in
\cite{Anikin} and, in more general form, in \cite{PSI} by using
the time-ordering $P$-exponential. It leads to the appearance of
additional diagrams in some electromagnetic processes.
However, they do not appear in the anomalous processes like
$\pi^0\to\gamma\gamma$ or $\gamma\to 3\pi$.

The coupling constant $g_\pi$ is determined from the so-called
{\it compositeness condition} or, equivalently, from the normalization
of the electromagnetic form factor to one at the origin.
Its value as well the value of pion weak coupling constant are defined by

\begin{eqnarray}\label{coupling}
\left({3g_\pi^2\over 4\pi^2}\right)^{-1}&\simeq&
{1\over 4}\int\limits_0^\infty duuf^2(-u){(3m_q^2+2u)\over (m^2_q+u)^3}.
\nonumber\\
& &\\
f_\pi
&\simeq &{3g_\pi\over 4\pi^2}m_q \int\limits_0^\infty duuf(-u)
{1\over (m_q^2+u)^2}.
\nonumber
\end{eqnarray}
Here, the pion mass is neglected since its magnitude is much less than
the cut-off parameter (see, \cite{Anikin}).

The form factor which defines the two-photon  decay of the pion
with the pion and photon momenta being off mass shell is written in the form

\begin{eqnarray}\label{PGG}
G_{\pi\gamma\gamma}(p^2,q_1^2,q_2^2)&=&
m_q\cdot {g_\pi\over 2\sqrt{2}\pi^2} R_{\gamma\pi}(p^2,q_1^2,q_2^2)
\\
&&\nonumber\\
&&\nonumber\\
\label{int}
R_{\gamma\pi}(p^2,q_1^2,q_2^2)&=&\int{d^4k\over\pi^2i}{f(k^2)\over D_0}
\end{eqnarray}
$
D_0=[m_q^2-(k+p/2)^2][m_q^2-(k-p/2)^2][m^2_q-(k+(q_1-q_2)/2)^2].
$
Using the Feynman $\alpha$-parametrization for $1/D_0$ one finds

\begin{equation}
R_{\gamma\pi}(p^2,q_1^2,q_2^2)=
2\int d^3\alpha\delta\left(1-\sum\limits_{i=1}^3\alpha_i\right)
\int{d^4k\over\pi^2i}{f(k^2)\over \biggl[m_q^2-D_1-(k+r)^2\biggr]^3}
\end{equation}
where
$D_1=\alpha_1\alpha_2 p^2+\alpha_1\alpha_3 q_1^2+\alpha_2\alpha_3 q_2^2$
and $r=q_2(1-2\alpha_2)/2-q_1(1-2\alpha_1)/2$.
Then we use the Cauchy theorem for the function

$$f(k^2)=\oint \frac{dz}{2\pi i} \frac{f(z)}{z-k^2},$$
and, again, the Feynman $\alpha$-parametrization.
Performing the integration over $k$ gives

$$
R_{\gamma\pi}(p^2,q_1^2,q_2^2)=\int\limits_0^\infty \! dt
\biggl(\frac{t}{1+t}\biggr)^2 \!
\int \! d^3\alpha \delta\left(1-\sum\limits_{i=1}^3 \alpha_i\right)
\biggl\{-f^\prime\biggl[-t(m^2_q-D_1)+\frac{t}{1+t}r^2\biggr]\biggr\}
$$
where $r^2=-D_1+p^2(1-2\alpha_3)/4+(q_1^2+q_2^2)\alpha_3/2$.
Note that this representation is valid for any function $f(k^2)$ and
convenient for numerical calculations.

The two-photon decay coupling constant is obtained from Eq.~(\ref{PGG})
for both $q_1^2$ and $q_2^2$ being equal to zero

\begin{equation}
g_{\pi\gamma\gamma}=G_{\pi\gamma\gamma}(m_\pi^2,0,0)
\simeq m_q{g_\pi\over 2\sqrt{2}\pi^2}
\int\limits_0^\infty duuf(-u){1\over (m_q^2+u)^3}.
\end{equation}

If we use the Gaussian form for the vertex function
$f(k^2)=\exp\{k^2/\Lambda^2\}$ the best fit is found to be
$f_\pi=132$ MeV ($f_\pi^{\rm expt}=132$ MeV) and
$g_{\pi\gamma\gamma}=0.261$ GeV$^{-1}$
($g_{\pi\gamma\gamma}^{\rm expt}=0.276$ GeV$^{-1}$)
with the adjustable parameters being equal to $\Lambda=1$ GeV
and $m_q=237$ MeV.

\section{The transition form factor}
\baselineskip 20pt

Defining the transition form factor by

\begin{equation}\label{ff}
F_{\gamma\pi}(Q^2)=e^2G_{\pi\gamma\gamma}(m_\pi^2,-Q^2,0)\simeq
e^2{g_\pi\over 2\sqrt{2}\pi^2} {m_q\over\Lambda^2}
R_{\gamma\pi}(Q^2)
\end{equation}
Then the decay width is written as

\begin{equation}\label{width}
\Gamma(\pi^0\to\gamma\gamma)=\frac{m_\pi^3}{64\pi}F_{\gamma\pi}^2(0).
\end{equation}

Our results for $(m_\pi^3/64\pi)F_{\gamma\pi}^2(Q^2)$ are shown in Fig.1
comparing with the experimental data and the monopole fit

\begin{equation}\label{fit}
F_{\gamma\pi}^{\rm fit}(Q^2)=
{e^2g_{\pi\gamma\gamma}\over 1+Q^2/\Lambda_\pi^2} \hspace{1cm}
\Lambda_\pi=0.6 \; {\rm GeV}.
\end{equation}

One can see that the form factor is close to the available experimental data.
 The radius for $\gamma^{\star}\pi^0\rightarrow\gamma$ transition is defined
by
\begin{equation}
<r^2_{\gamma\pi}>=-6\frac{F^{\prime}_{\gamma\pi}(0)}{F_{\gamma\pi}(0)}
\end{equation}
where
\begin{equation}
F_{\gamma\pi}(0)=\int\limits_{0}^{\infty}duu\frac{f(-u)}{(m_q^2+u)^3}
\hspace{1cm}
F^{\prime}_{\gamma\pi}(0)=-\frac{m_q^2}{2}\int\limits_{0}^{\infty}
duu\frac{f(-u)}{(m_q^2+u)^5}.
\end{equation}
With this definition one finds $r_{\gamma\pi}$=0.65 fm which is the
experimental value having an error of 0.03 fm.

We compare our result for the function
$Q^2F_{\gamma\pi}(Q^2)/F_{\gamma\pi}(0)$ with the Brodsky-Lepage limit
$4\pi^2 f^2_\pi\approx 0.68$ GeV$^2$ in the Fig.2.
One can see that this function reaches smoothly the Brodsky-Lepage limit
at $Q^2\approx 10$ GeV$^2$.

Summarizing we have calculated the pion weak decay constant, the two-photon
decay width,
as well as the form factor of the $\gamma^*\pi^0\to\gamma$-transition.
The two adjustable parameters,
the range parameter $\Lambda$ appearing in the Gaussian, and
the constituent quark mass $m_q$, have been fixed by
fitting the experimental data for the pion decay constants.
We find that the $\gamma\pi$ - form factor is close to
the available experimental data and
smoothly reaches the Brodsky-Lepage limit
at $Q^2\approx 10$ GeV$^2$.

\section{Acknowledgements}

We acknowledge helpful discussions with M.P. Locher during our stay
at PSI. This work was supported in part by the INTAS Grant 94-739
and the Russian Fund of Fundamental Research under contract 96-02-17435-a.

\newpage

\begin{center}
List of figures
\end{center}

\vspace{1cm}

\noindent
Fig.1. Our calculated form of
$m_\pi^3 F_{\gamma\pi}^2(Q^2)/64\pi$ (dashed line) compared
with experimental data and their fit (solid line) \cite{CELLO,CLEO}.\\
Fig.2. Our calculated form of $Q^2F_{\gamma\pi}(Q^2)/F_{\gamma\pi}(0)$
(dashed line) compared  with experimental fit (solid line) and
the Brodsky-Lepage limit ($\approx$ 0.68 GeV$^2$) \cite{Brodsky}.

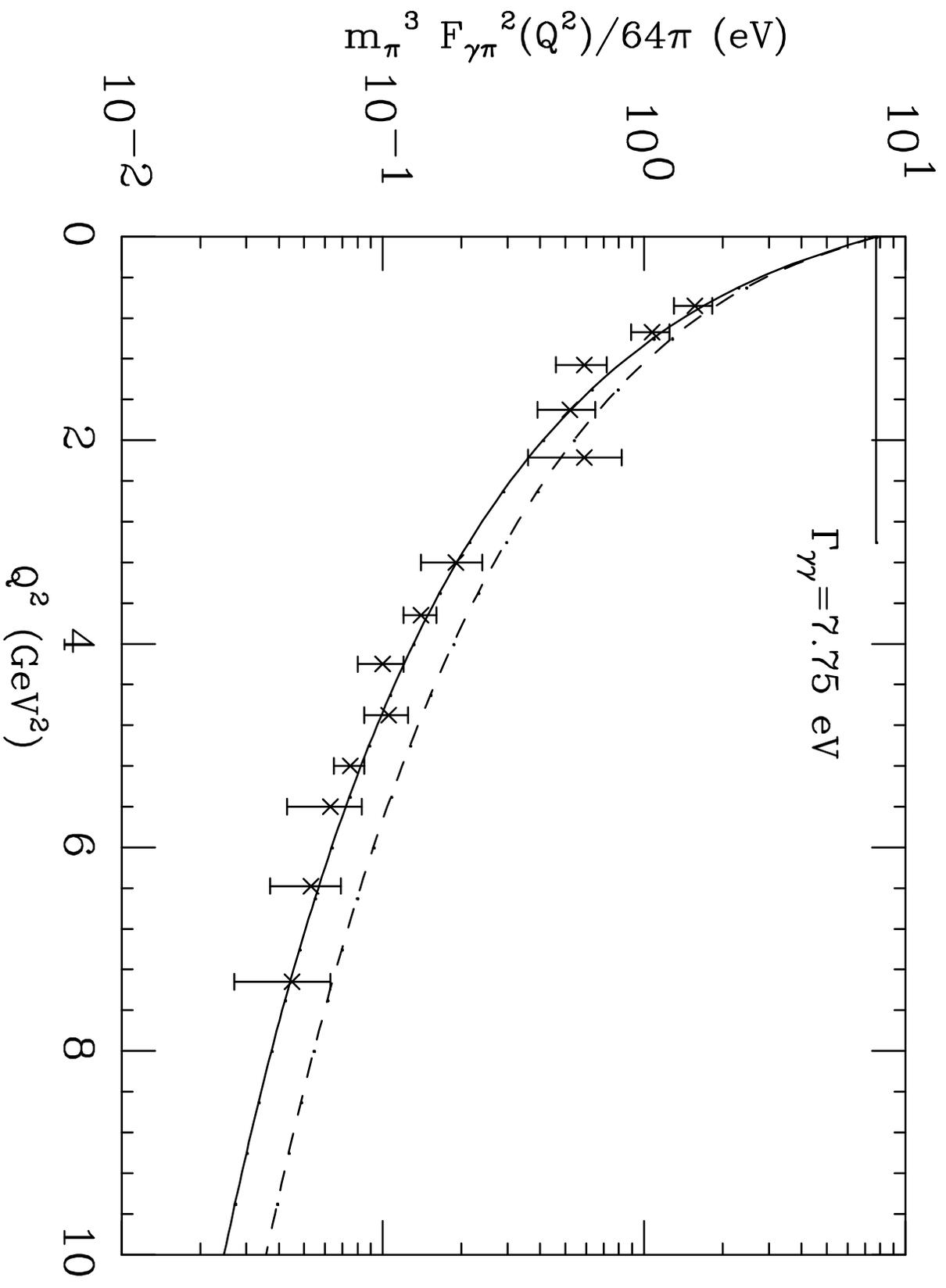
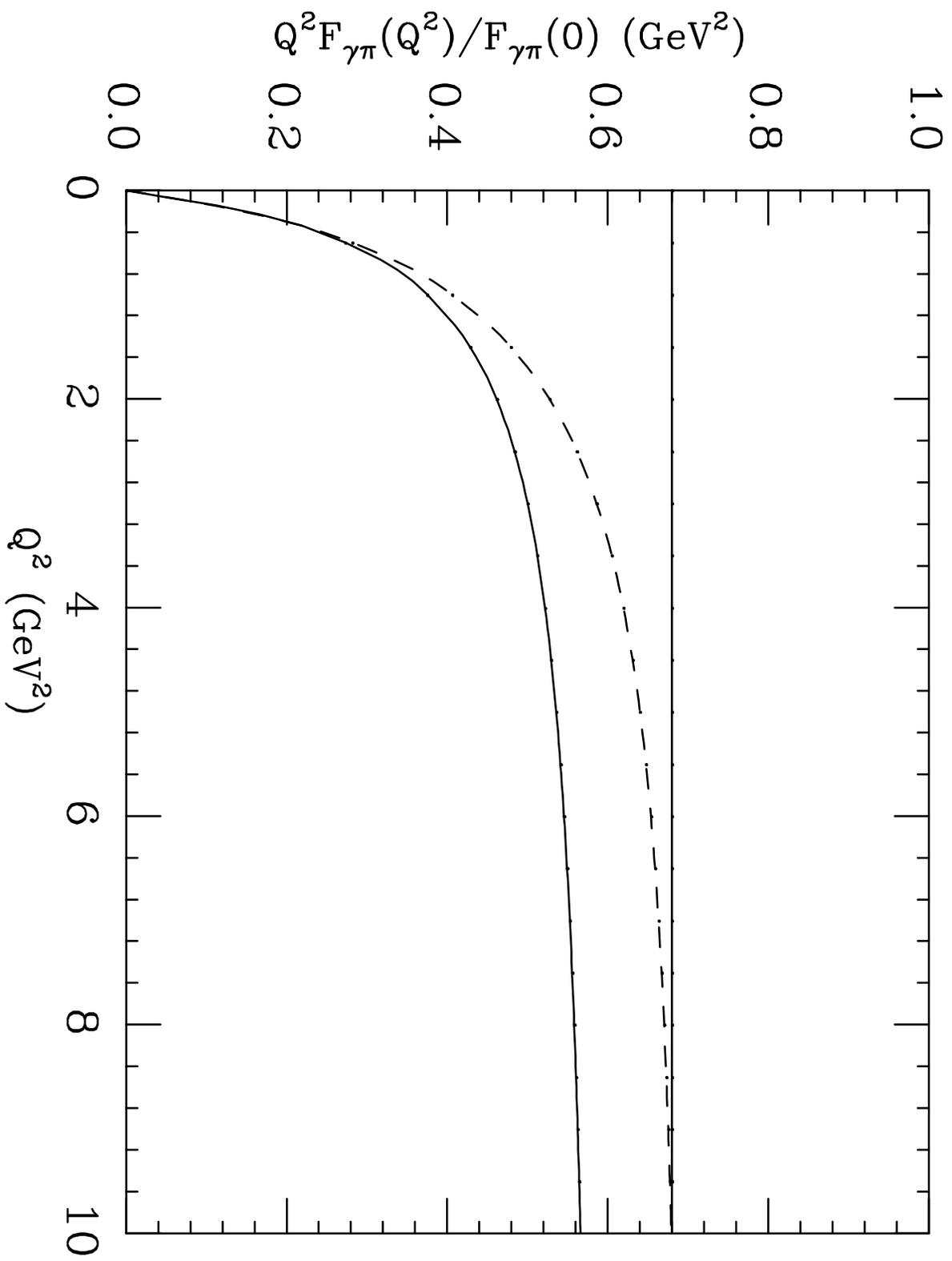
\end{document}